\documentclass[12pt]{iopart}
\usepackage{iopams}

\newtheorem{theorem}{Theorem}[section]
\newtheorem{proposition}[theorem]{Proposition}
\newtheorem{lemma}[theorem]{Lemma}

\newtheorem{definition}[theorem]{Definition}

\newlength{\vshift}
\newlength{\hshift}
\usepackage{amssymb,amsopn}

\begin{document}
\begin{flushright}
ICMPA-MPA/008/2012
\end{flushright}
\title{$(q;l,\lambda)$-deformed Heisenberg algebra: representations, special functions and coherent states quantization}
\author{Mahouton Norbert Hounkonnou,  Sama Arjika and Ezinvi Baloitcha}
\address{International Chair of Mathematical Physics
and Applications (ICMPA-UNESCO Chair), University of
Abomey-Calavi, 072 B. P.: 50 Cotonou, Republic of Benin}
\eads{\mailto{norbert.hounkonnou@cipma.uac.bj},
\mailto{rjksama2008@gmail.com},
\mailto{ezinvi.baloitcha@cipma.uac.bj}}

\begin{abstract}
This paper addresses
  a new characterization of 
Sudarshan's diagonal 
representation of the density matrix elements $\rho(z',z)$, derived
from   $(q;l,\lambda)-$deformed boson coherent states.
The induced $\rho(z',z)$ self-reproducing property 
with the  associated  self-reproducing kernel $K(z',z)$ is computed and analyzed. 
An explicit construction of novel classes of generalized continuous  
$(q;l,\lambda)-$Hermite polynomials is provided with the corresponding  recursion relations and 
 exact  resolution of the moment problems giving their orthogonality weight functions. Besides, the Berezin-Klauder-Toeplitz  quantization of classical phase space observables and  relevant normal and anti-normal forms    are investigated and  discussed. 
\end{abstract}

 \today

\section{Introduction}
The Heisenberg algebra,  generated by the  identity operator $\mathbf{1}$ and two mutually adjoint operators,  $b$ and   its Hermitian conjugate $b^\dag$ (also  called annihilation and  creation operators in Physics literature),   satisfying the commutation relations
\begin{eqnarray}\label{Heise}
 [b,\;b^\dagger]=\mathbf{1},\qquad [b,\;\mathbf{1}]=0=[b^\dagger,\;\mathbf{1}],
\end{eqnarray}
where  $[A, \;B]:= AB - BA,$
plays a central role in the
investigation of physical systems and  in mathematics. 

Defining the operator  $N:=b^\dag b,$  known as  the  {\it number operator}, the commutation relations (\ref{Heise})
induce the two following  properties:
\begin{eqnarray}\label{Harmonic2}
 [N,\;b]=-b\quad\mbox{and}\quad [N,\;b^\dag]=b^\dag.
\end{eqnarray}
Let $\mathcal{F}$  be a Fock space and $\{|n\rangle\;|\;n\in \mathbb{N}\cup \{0\}\}$ be its orthonormal  basis. The actions of $b$, $b^\dag$ and $N$ on  $\mathcal{F}$ are given by 
\begin{eqnarray}\label{Harmonic2b}
 b|n\rangle= \sqrt{n}|n-1\rangle,\;\;b^\dag|n\rangle=\sqrt{n+1}|n+n\rangle,\;\mbox{ and }\; N|n\rangle=n|n\rangle
\end{eqnarray}
where  $|0\rangle$ is a normalized vacuum:
\begin{eqnarray}
 b|0\rangle= 0,\qquad \langle 0|0\rangle=1.
\end{eqnarray}
From (\ref{Harmonic2b}) the states $|n\rangle$ for $n\ge1$ are built as follows:
\begin{eqnarray}
 |n\rangle=\frac{1}{\sqrt{n!}}(b^\dagger)^n|0\rangle,\;\; n= 1,\;2,\;\cdots
\end{eqnarray}
satisfying the orthogonality and completeness conditions:
\begin{eqnarray}
 \langle m|n\rangle=\delta_{m,n}, \;\quad \sum_{n=0}^\infty |n\rangle \langle n|= \mathbf{1}.
\end{eqnarray}

The generalization of  the canonical commutation relations  (\ref{Heise}) was suggested long before the discovery of quantum groups, by Heisenberg to achieve the regularization for  nonlinear spinor field theory. The issue was considered as small additions to the canonical commutations relations \cite{Dancoff,Tamm}. 
Snyder, investigating the infrared catastrophe of soft photons in the Compton scattering, raised this issue   and built a non-commutative Lorentz invariant space-time where the non-commutativity of space operators is proportional to non-linear combinations of phase space operators \cite{Snyder}. 
Further,
the deformation of the harmonic oscillator algebra  whose applications in physics are presently 
rather technical but nonetheless very promising \cite{zhang}, possesses an important and useful representation theory
  in  connection to that of their classical limit algebra. The investigation under the form $a a^\dag-q a^\dag a={\bf 1}, \,q>1$ of the one-parameter deformed
Heisenberg  algebras in theoretical physics originated from
 the study of the dual resonance models of strong
interactions \cite{arik}. 

From the other side, there are some hopes that,
 in physical studies of nonlinear phenomena, the deformed oscillator can play the same role   as the usual boson oscillator in
 non-relativistic quantum mechanics. This could explain why various quantum deformations of boson oscillator commutation relations
  have attracted a great attention during the last few years (see 
\cite{Balo} and references therein).
 This might be also due to the fact that there exist correspondences between quantum groups, quantum algebras, 
statistical mechanics, quantum field theory, conformal 
field theory, quantum and nonlinear optics  and non commutative geometry, etc. 
 Furthermore, such a connection is extended to
coherent states  deducible from
the  study of quantum groups and, therefore, from 
 the deformation of Heisenberg algebra.  
Recently  \cite{dez},
 a deformation of
the Heisenberg algebra  by a set of parameters 
was introduced
with  a new family of 
generalized coherent states respecting the Gazeau-Klauder's criteria. 

 Besides, Parthasarathy  and  Sridhar studied 
a $q-$analogue of the diagonal  representation of the density matrix using the 
$q-$boson coherent states and gave the generalization of 
the self-reproducing property of  density matrix elements $\rho(z',z)$ and the associated kernel $K(z',z)$ (see \cite{part1,part2} for more details). 

The present work addresses a new characterization of 
Sudarshan's diagonal 
representation of the density matrix \cite{sudar},  derived
from the constructed  $(q;l,\lambda)-$deformed boson coherent states.
The $(q;l,\lambda)-$generalization of the self-reproducing property of  density matrix elements
$\rho(z',z)$ and associated  self-reproducing kernel $K(z',z)$ are computed and analyzed.  New families of generalized Hermite polynomials  associated with the position and momentum operators  as well as the main relevant operator properties are investigated and discussed.

The paper is organized as follows.  In section 2, we first give  main useful definitions and  results on the $(q;l,\lambda)-$deformed oscillator algebra and related coherent states. Then  we  deduce resulting new features  used in the sequel.  In section 3,
 we provide an explicit construction, including the recursion relation of generalized continuous  
$(q;l,\lambda)-$Hermite polynomials 
generated by polynomial expansion of  the deformed position and momentum operators in associated Fock space basis.  Particular classes of deformed Hermite polynomials are deduced  with explicit  
weight functions. In the section 4,  
diagonal representation of density matrix using the $(q;l,\lambda)-$CS is computed. Reproducing kernel $K(z,\zeta)$  and its properties are investigated. In the section 5, matrix elements of normal and anti normal forms  and  mean operator values  are determined.  The Berezin-Klauder-Toeplitz  quantization (also called CS quantization) of classical phase space observables is performed in section 6.  Furthermore, the  angle and   time evolution  operators and  semi-classical phase space trajectories are discussed. 
Finally, we end by a conclusion in the section 7, followed by appendices on  some computational details.
\section{On the $(q;l,\lambda)-$deformed oscillator algebra} 
In this section, for the clarity of our development, we first  recall main definitions and 
 results on the $(q;l,\lambda)-$deformed oscillator algebra \cite{dez}, and  then deduce resulting new features  used in the sequel.
\begin{definition}
The $(q;l,\lambda)-$deformed oscillator algebra is defined as the associative 
algebra generated by the operators $\{ {\bf 1},\;a,\;a^\dag,\;N\}$ 
satisfying the commutation relations \cite{dez}
\begin{eqnarray}
\label{rela}
\fl 
\qquad\quad aa^\dag-a^\dag a=l^{2}q^{\lambda-N-1},\quad [N,a^\dag]=a^\dag,\quad [N,a]=-a,\quad 
\varphi(n)=l^2q^\lambda\frac{1-q^{-n}}{q-1},
\end{eqnarray} 
where $\varphi$ is the structure function, $l$ and $ \lambda$ are complex numbers with $l\neq 0$ and 
$q>0.$
\end{definition}
 This algebra carries  out a Hopf algebra structure (see Appendix A).
The operator products  $a a^\dag$ and $a^\dag a$ are obtained from \eref{rela} and are given by
\begin{eqnarray}
\label{taz}
a a^\dag=l^2q^\lambda\frac{1-q^{-N-1}}{q-1},\quad 
a^\dag a= l^2q^\lambda\frac{1-q^{-N}}{q-1}.
\end{eqnarray}
\begin{proposition}
\cite{dez}
The orthonormalized basis of the Fock space $\mathcal{F}$ is given  by
\begin{eqnarray}
\label{ket}
|n\rangle:=
\frac{q^{\frac{1}{2}({}^n_2)}}{\sqrt{\gamma^{n}(q;q)_n}}a^{\dagger n}|0\rangle,\quad  n=0,1,2,\ldots
\end{eqnarray}
where $\gamma=l^2q^{\lambda-1}/(1-q);$  the $q-$shifted factorial $(q;q)_n$ is defined as:  $(z;q)_n:=\prod_{k=0}^{n-1}(1-zq^k), \;n=1,2,\ldots$ with $(z;q)_0:=1$  by convention.
The states \eref{ket} satisfy the orthogonality and completeness conditions
\begin{eqnarray}
 \langle m|n\rangle=\delta_{m,n}, \;\; \sum\limits_{n=0}^\infty |n\rangle\, \langle n|= \mathbf{1}.
\end{eqnarray}
Moreover, the  actions of the operators $a, a^\dag, aa^\dagger, a^\dagger a$ and $N$ on (\ref{ket})
are given by
\begin{eqnarray}
\label{actions}
a|n\rangle=\sqrt{\varphi(n)}|n-1\rangle,\quad 
a^\dag|n\rangle=\sqrt{\varphi(n+1)}|n+1\rangle,\\
a a^\dagger|n\rangle=\varphi(n+1)|n\rangle,\;\;
a^\dagger a|n\rangle =\varphi(n)|n\rangle,\;\;
N|n\rangle=n|n\rangle.
\end{eqnarray}
\end{proposition}
{\bf Proof.}  See \cite{dez}. $\square$
\begin{definition}
The  $(q;l,\lambda)-$Jackson's differential operator ${_q}\partial_y^{l,\lambda}$ acting on the space of analytic functions is defined as \cite{dez}
\begin{eqnarray}
\label{opera}
{_q}\partial_y^{l,\lambda}f(y):=l^2q^{\lambda}\frac {f(y)-f(q^{-1}y)}{(q-1)y}.
\end{eqnarray}
\end{definition}

The deformed $(q;l,\lambda)-$coherent states (CS) associated with the algebra (\ref{rela}) are constructed in \cite{dez}:
\begin{eqnarray}
\label{des}
|z\rangle_{l,\lambda}=\mathcal{N}_{l,\lambda}^{-1/2}(|z|^2)\sum_{n=0}^\infty\frac{q^{n(n-1)/4}z^n}{\sqrt{\gamma^n(q;q)_n}}|n\rangle, 
\quad z\in\mathbf{D}_{l,\lambda},
\end{eqnarray}
where 
\begin{eqnarray}
\label{dess}
\mathcal{N}_{l,\lambda}(t)=\sum_{n=0}^\infty\frac{q^{({}^n_2)}t^n}{\gamma^n(q;q)_n} ,
\end{eqnarray}
and 
\begin{eqnarray}
\label{radius}
\mathbf{D}_{l,\lambda}=\{z\in\mathbb{C}: |z|<R_{l,\lambda}\}, \quad  \quad R_{l,\lambda}=\left\{\begin{array}{ll}\infty& \mbox{ if } \;0<q<1\\
\frac{l^2q^{\lambda}}{q-1} &  \mbox{ if } \; q> 1,\end{array}\right.
\end{eqnarray}
with $(z;q)_\infty :=\prod_{k=0}^\infty(1-zq^k).$ $R_{l,\lambda}$ is the convergence radius of 
the series $\mathcal{N}_{l,\lambda}(t)$ which
 is a holomorphic function
 with simple  zeros  at $x_k=l^2q^{\lambda-1-k}/(q-1),\;k=0,1,...$

The CS (\ref{des})  are not orthogonal as we can see from the product of two CS $|z\rangle_{l,\lambda}$ and
 $|z'\rangle_{l,\lambda}$
\begin{eqnarray}
\label{prod}
_{l,\lambda}\langle z'|z\rangle_{l,\lambda}
=\left[\mathcal{N}_{l,\lambda}(|z'|^2)\mathcal{N}_{l,\lambda}(|z|^2)\right]^{-1/2}
\sum_{n=0}^\infty\frac{q^{({}^n_2)}
}{(q;q)_n}\left(\frac{\bar{z}'z}{\gamma}\right)^n .
\end{eqnarray}
 Besides,  it is proved in \cite{dez} that they
solve the identity, i.e,
\begin{eqnarray}
\label{resolv}
\int_{
\mathbf{D}_{l,\lambda}}d\mu_{l,\lambda}(\bar{z},z)|z\rangle_{l,\lambda}\,{_{l,\lambda}\langle z}|={\bf 1},
\end{eqnarray}
where
\begin{eqnarray}
\label{ress}
\fl
\quad d\mu_{l,\lambda}(\bar{z},z)=
\left\{\begin{array}{ll}\frac{1}{\eta \ln q^{-1}}\frac{\mathcal{N}_{l,\lambda}(\bar{z}z)}{\mathcal{N}_{l,\lambda}(\bar{z}zq^{-1})}\frac{d^2z}{\pi}, &  0 < q < 1\\
\frac{1}{2\pi}\frac{d_q^{l,\lambda}x\,d\theta}{1+ \frac{x}{\eta}}, & q>1,\; 0 < x=|z|^2 < \frac{l^2q^\lambda}{q-1},\;\theta=arg(z),\end{array}
\right.
\end{eqnarray}
 with $\eta=l^2q^\lambda/(1-q).$
\section{Deformed Hermite polynomials  associated with the position 
and momentum operators}
In \cite{dez}, it is proved  that the deficiency indices of the position and momentum operators,
$Q=(a^\dag+a)/\sqrt 2$ and $P=i(a^\dag-a)/\sqrt 2,$ \;is \;$(1,1)$. Therefore, they are no longer 
essentially self-adjoint
but have each a one-parameter family of self-adjoint extensions instead. 
In this case, the deficiency subspaces $N_x$,
$Im(x) \ne 0,$ are one-dimensional. Associate  now these operators  to the generalized vectors 
\begin{eqnarray}
\label{generalize}
|x \rangle:=\sum_0^\infty q_n (x)|n\rangle,\\
\label{generalized}
|p \rangle:=\sum_0^\infty p_n (p)|n\rangle,
\end{eqnarray}
respectively,
 such that their actions 
are  realized as follows:
\begin{eqnarray}
\label{u}
Q|x\rangle=x|x\rangle,\quad P|p\rangle=p|p\rangle
\end{eqnarray} 
and analyze their various relevant representations.
\subsection{ In the position representation: $(q;l,\lambda)-$deformed Hermite polynomials}
Here two cases deserve examination depending on the $q-$value range.\\
$\bullet$ {\bf Case 1:}  $0 <q <1$\\
By using the equations (\ref{actions}), (\ref{generalize}) 
and (\ref{u}), we readily prove  the following recurrence relation obeyed by the Fock space basis coefficients $q_n:$
\begin{eqnarray}
\label{set}
\fl
\quad\sqrt{2(1-q)}xq_n(x)=(l^2q^{\lambda-n-1}(1-q^{n+1}))^{1/2}q_{n+1}(x)+
(l^2q^{\lambda-n}(1-q^{n}))^{1/2}q_{n-1}(x),
\end{eqnarray}
imposing the initial conditions $q_{-1}(x):= 0,\;q_0(x):= 1.$
By setting $2^{1/2}y=(1-q)^{1/2}x$ and $\psi_n(y|q)=q_{n}(\sqrt{2(1-q)^{-1}}y)$, the equation  (\ref{set}) can be re-expressed as
\begin{eqnarray}
\label{sett}
\fl
\qquad 2y\psi_n(y|q)=(l^2q^{\lambda-n-1}(1-q^{n+1}))^{1/2}\psi_{n+1}(y|q)+
(l^2q^{\lambda-n}(1-q^{n}))^{1/2}\psi_{n-1}(y|q).
\end{eqnarray}
Putting now $\psi_n(x|q)=(l^2q^{\lambda})^{-\frac{n}{2}}q^{\frac{n(n+1)}{4}}(q;q)_n^{-1/2}h_n(x;l,\lambda|q) $ transforms the formula (\ref{sett})
into the new recursive relation  
\begin{eqnarray}
\label{setrt}
2xh_n(x;l,\lambda|q)=h_{n+1}(x;l,\lambda|q)+l^2q^\lambda(q^{-n}-1)
h_{n-1}(x;l,\lambda|q)
\end{eqnarray}
defining  a novel family of $(q;l,\lambda)-$deformed Hermite polynomials, i.e.  $\{h_n(x;l,\lambda|q),\;n=0,1,2,....\}.$
The Fock space basis  vector coefficients $q_n(x)$ giving the  eigenvectors of the operator $Q=(a^\dag+a)/\sqrt 2$ by the expansion  (\ref{generalize}) are then explicitly given by
\begin{eqnarray} 
\label{100} 
q_n(x)=(l^2q^{\lambda})^{-\frac{n}{2}}q^{n(n+1)/4}(q;q)_n^{-1/2}h_n(\sqrt{2^{-1}(1-q)}x;l,\lambda|q).
\end{eqnarray}
 
In the particular case, when $\lambda= 0$ and $l=1$,  (\ref{100}) provides  the recursive relation (1.7) obtained in \cite{askey} (replacing  $q^{-1}$  by $q$, $q>1$) for the continuous $q-$Hermite polynomials 
 $h_n(x|q)=i^{-n}H_n(ix|q)$ for $q>1$.
In this case, the expansion coefficients $q_n$ reduce to
\begin{eqnarray}
\label{10t0}
q_n(x)=q^{n(n+1)/4}(q;q)_n^{-1/2}h_n(\sqrt{2^{-1}(q-1)}x|q)
\end{eqnarray}
satisfying 
the orthogonality relation  \cite{askey} 
\begin{eqnarray}
\label{ort}
\int_{-\infty}^\infty q_m(b^{-1}\sinh u)q_n(b^{-1}\sinh u)d\mu(u)=\delta_{n,m},
\end{eqnarray}
where $b=\sqrt{2/(q-1)},$
\begin{eqnarray}
\label{begu}
h_n(\sinh u|q)=\sum_{k=0}^n(-1)^kq^{k(k-n)}\Big[\begin{array}{c}n\\k\end{array}\Big]_qe^{(n-2k)u},
\end{eqnarray}
 and the measure $d\mu(u)$ is given by
\begin{eqnarray}
\label{mea}
d\mu(u)=\frac{du}{(q;q)_\infty\ln q^{-1}\prod_{k=1}^{\infty}(1+2\cosh 2u\,q^k+q^{2k})}.
\end{eqnarray}
$\bullet$ {\bf Case 2:} $q >1$\\
From the same equations \eref{actions}, (\ref{generalize}) 
and (\ref{u}), we arrive at the following recursive relation
\begin{eqnarray}
\label{set}
\fl
\qquad \sqrt{2(q-1)}xq_n(x)=(l^2q^{\lambda}(1-q^{-n-1}))^{1/2}q_{n+1}(x)+
(l^2q^{\lambda}(1-q^{-n}))^{1/2}q_{n-1}(x),
\end{eqnarray}
with the initial conditions $q_{-1}(x):= 0,\;q_0(x):= 1.$
Performing the same development as above, 
this recursive relation is re-arranged in the form
\begin{eqnarray}
\label{edsetrt}
2x\hat{h}_n(x;l,\lambda|q)=\hat{h}_{n+1}(x;l,\lambda|q)+l^2q^\lambda(1-q^{-n})
\hat{h}_{n-1}(x;l,\lambda|q),
\end{eqnarray}
where $\{\hat{h}_n(x;l,\lambda|q),\;n=0,1,2,....\}$ constitutes a new family of $(q;l,\lambda)-$deformed Hermite polynomials determining 
 the coefficients $q_n(x)$ (\ref{generalize}) as follows:
\begin{eqnarray} 
\label{10d0} 
q_n(x)=(l^2q^{\lambda})^{-\frac{n}{2}}(q^{-1};q^{-1})_n^{-1/2}\hat{h}_n(\sqrt{2^{-1}(q-1)}x;l,\lambda|q).
\end{eqnarray}
reverting a simpler form for specific values
 $\lambda= 0$ and $l=1,$ i.e.
\begin{eqnarray} 
\label{ferr} 
q_n(x)=(q^{-1};q^{-1})_n^{-1/2}\hat{h}_n(\sqrt{2^{-1}(q-1)}x|q),
\end{eqnarray}
where $\hat{h}_n(\sqrt{2^{-1}(q-1)}x|q)$ are the continuous $q-$Hermite polynomials taking 
with the base $q^{-1}$ \cite{sang}. The polynomials \eref{ferr} satisfy the following orthogonality relation
\begin{eqnarray}
\label{ororrt}
\int_{-c}^cq_m(x)q_n(x)\hat{\mu}(x)dx=\delta_{n,m},
\end{eqnarray}
where $c=\sqrt{2/(q-1)}$ and the measure $\hat{\mu}(x)$ is given by
\begin{eqnarray}
\label{mea2}
\hat{\mu}(x)=\;\;\frac{(q^{-1};q^{-1})_\infty}{\pi\sqrt{q-1} }
\frac{\prod_{k=0}^{\infty}(1-2((q-1)x^2-1)
q^{-k}+q^{-2k})}{\sqrt{2-(q-1)x^2}}.
\end{eqnarray}
\subsection{ In the momentum representation: $(q;l,\lambda)-$deformed Hermite polynomials}
Two $q-$ value situations merit also to be examined in this section.\\
$\bullet$ {\bf Case 1:} $0 <q <1$\\
The above equations  (\ref{actions}), \eref{generalized}
and (\ref{u}) result in the following recursion relation:
\begin{eqnarray}
\label{eset}
\fl
\;-\sqrt{2(1-q)}pp_n(p)&=&i(l^2q^{\lambda-n-1}(1-q^{n+1}))^{1/2}p_{n+1}(p)
-
i(l^2q^{\lambda-n}(1-q^{n}))^{1/2}p_{n-1}(p),
\end{eqnarray}
with the initial conditions $p_{-1}(p):= 0,\;p_0(p):= 1.$

Similarly to the case in the position representation, 
setting $2^{1/2}y=(1-q)^{1/2}p$, $\psi_n(y|q)=
p_{n}(\sqrt{2/(1-q)}y)$ and 
$\psi_n(y|q)=i^n(l^2q^{\lambda})^{-\frac{n}{2}}
q^{n(n+1)/4}(q;q)_n^{-1/2}\chi_n(y;l,\lambda|q)$ with $i=\sqrt{-1}$ 
leads to  the three term 
recursion relation satisfied by $\chi_n(y;l,\lambda|q)$
\begin{eqnarray}
\label{srrt}
2y\chi_n(y;l,\lambda|q)=
\chi_{n+1}(y;l,\lambda|q)+l^2q^{\lambda}(q^{-n}-1)
\chi_{n-1}(y;l,\lambda|q),
\end{eqnarray}
defining  a new family 
of $(q;l,\lambda)-$deformed polynomials, i.e.  $\{\chi_n(y;l,\lambda|q),\;n=0,1,2,...\}$.
The Fock space basis  vector coefficients $p_n(p)$  giving the  eigen-vectors of the operator  $P=i(a^\dag-a)/\sqrt 2$   by the expansion  (\ref{generalized}) are then explicitly given by
\begin{eqnarray}
\label{1ss00}
p_n(p)=(i^{-1}lq^{\lambda/2})^{-n}
q^{n(n+1)/4}(q;q)_n^{-1/2}
\chi_n(\sqrt{2^{-1}(1-q)}p;l,\lambda|q).
\end{eqnarray}

In the specific case of  $\lambda= 0$ and $l= i$,  (\ref{1ss00}) coincides with the recursion relation (1.7) in \cite{askey} (if $q^{-1}$ is replaced by $q$, $q>1$) for the continuous $q-$Hermite polynomials $\chi_n(p|q)=i^{-n}H_n(ip|q)$ when $q>1$.
In this case,
\begin{eqnarray}
\label{1dd0t0}
p_n(p)=q^{n(n+1)/4}(q;q)_n^{-1/2}\chi_n(\sqrt{2^{-1}(1-q)}p|q)
\end{eqnarray}
satisfying 
 the orthogonality relation \cite{askey}
\begin{eqnarray}
\label{orddt}
\int_{-\infty}^\infty p_m(b^{-1}\sinh v)p_n(b^{-1}\sinh v)d\nu(v)=\delta_{n,m},
\end{eqnarray}
where $b=\sqrt{2/(1-q)}$
 and the measure $d\nu(v)$ is given by
\begin{eqnarray}
\label{meea}
d\nu(v)=\frac{dv}{(q;q)_\infty\ln q^{-1}\prod_{k=1}^{\infty}(1+2\cosh 2v\,q^k+q^{2k})}.
\end{eqnarray}
$\bullet$ {\bf Case 2:} $q >1$\\
In this case,  the equations \eref{actions}, (\ref{generalized}) 
and (\ref{u})  also allow to produce the following recursion relation
\begin{eqnarray}
\label{set}
\fl
\quad -\sqrt{2(q-1)}pp_n(p)=i(l^2q^{\lambda}(1-q^{-n-1}))^{1/2}p_{n+1}(p)
-i(l^2q^{\lambda}(1-q^{-n}))^{1/2}p_{n-1}(p),
\end{eqnarray}
with the initial conditions $p_{-1}(p):= 0,\;p_0(p):= 1.$ Furthermore, following step by step the above development, we arrive at the relation
\begin{eqnarray}
\label{edsetrt}
2y\hat{\chi}_n(y;l,\lambda|q)=\hat{\chi}_{n+1}(y;l,\lambda|q)+l^2q^\lambda(1-q^{-n})
\hat{\chi}_{n-1}(y;l,\lambda|q),
\end{eqnarray}
determining a new family of $(q;l,\lambda)-$deformed Hermite polynomials, i.e. $\{\hat{\chi}_n(y;l,\lambda|q),\\
\;n=0,1,2,....\}$  giving
the coefficients $p_n(p):$ 
\begin{eqnarray} 
\label{10ddd0} 
p_n(p)=(i^{-1}lq^{\lambda/2})^{-\frac{n}{2}}(q^{-1};q^{-1})_n^{-1/2}\hat{\chi}_n(\sqrt{2^{-1}(q-1)}p;l,\lambda|q).
\end{eqnarray}

In the situation, when $\lambda= 0$ and $l=  i$,  (\ref{10ddd0}) simplifies to
\begin{eqnarray} 
\label{fer} 
p_n(p)=(q^{-1};q^{-1})_n^{-1/2}\hat{\chi}_n(\sqrt{2^{-1}(q-1)}p|q),
\end{eqnarray}
where $\hat{\chi}_n(\sqrt{2^{-1}(q-1)}p|q)$ are the continuous $q-$Hermite polynomials taking 
with the base $q^{-1}$ \cite{sang}, with    
the orthogonality relation
\begin{eqnarray}
\label{ororrt}
\int_{-c}^cp_m(p)p_n(p)\hat{\nu}(p)dp=\delta_{n,m},
\end{eqnarray}
where $c=\sqrt{2/(q-1)}$ and the measure $\hat{\nu}(p)$ is given by
\begin{eqnarray}
\label{mea2}
\hat{\nu}(p)=\;\;\frac{(q^{-1};q^{-1})_\infty}{\pi\sqrt{q-1} }
\frac{\prod_{k=0}^{\infty}(1-2((q-1)p^2-1)
q^{-k}+q^{-2k})}{\sqrt{2-(q-1)p^2}}.
\end{eqnarray}
\section{Diagonal representation }
\subsection{Diagonal representation of the density matrix}
 The definition of the density
 matrix uses the overcompleteness of  Fock space states as follows:
\begin{eqnarray}
\label{bel}
\rho:=\sum_{n,m=0}^\infty\rho(n,m) 
|n\rangle\langle m|.
\end{eqnarray}
 Letting  $z=re^{i\theta},\;0\leq \theta\leq 2\pi$
 and $0 < r <R_{l,\lambda}$ and making use of  
 (\ref{des})
amount  to
\begin{eqnarray}
\label{refr}
\mathcal{N}_{l,\lambda}(r^2)|re^{i\theta}
\rangle_{l,\lambda}\,_{l,\lambda} \langle 
re^{i\theta}|=\sum_{n,m=0}^\infty
\frac{r^{n+m}e^{i(n-m)\theta}q^{\frac{1}{2}({}^n_2)+
\frac{1}{2}({}^m_{\;2})}}{\sqrt{\gamma^{n+m}(q;q)_n(q;q)_m}}
|n\rangle\langle m|.
\end{eqnarray}
Now  by multiplying (\ref{refr}) by $e^{is\theta}$  and performing the  integral with respect to the angle $\theta$
 in the Lebesgue sense, 
 the $r-$integration being  the "$q; l, \lambda-$integration"
we obtain

\begin{eqnarray}
\label{trefr}
\fl
\qquad\;\int_{0}^{2\pi}\frac{d\theta}{2\pi}
\mathcal{N}_{l,\lambda}(r^2)e^{is\theta}|re^{i\theta}
\rangle_{l,\lambda}\,_{l,\lambda} \langle re^{i\theta}|
=\sum_{n,m=0}^\infty\frac{r^{n+m}q^{\frac{1}{2}({}^n_2)+
\frac{1}{2}({}^m_{\;2})}}{\sqrt{\gamma^{n+m}(q;q)_n(q;q)_m}}
|n\rangle\langle m|\delta_{s,m-n}.
\end{eqnarray}
By applying $p-$times the operator $_{q}\partial_r^{l,\lambda}$ 
on the two sides of (\ref{trefr})
and  evaluating the result 
at $r=0$, only  
the term
$n+m-p=0$ 
 survives in the right-hand side. It results
\begin{eqnarray}
\label{treefr}
\fl 
\Bigg\{\Big({_q}\partial_r^{l,\lambda}\Big)^p
\int_{0}^{2\pi}\frac{d\theta}{2\pi}
\mathcal{N}_{l,\lambda}(r^2)e^{is\theta}
|re^{i\theta}\rangle_{l,\lambda}\,_{l,\lambda} \langle re^{i\theta}|\Bigg\}_{r=0}=\cr
\sum_{n,m=0}^\infty\frac{q^{\frac{1}{2}({}^n_2)+
\frac{1}{2}({}^m_{\;2})+({}^p_2)}(q;q)_{n+m}}
{\sqrt{\gamma^{n+m}(q;q)_n(q;q)_m}(q;q)_{n+m-p}}
\Bigg(\frac{l^2q^{\lambda-n-m}}{1-q}\Bigg)^p
|n\rangle\langle m|
\delta_{s,m-n}\delta_{p,m+n},\cr
\end{eqnarray}
furnishing
\begin{eqnarray}
\label{treeefr}
|n\rangle
\langle m|&=&\Bigg(\frac{q^{({}^{n+m}_{\;\;\;2})+nm}}{ \gamma^{n+m}(q^{1+n};q)_m(q^{1+m};q)_n}\Bigg)^{\frac{1}{2}}\cr
&\times&\Bigg\{\Big({_q}\partial_r^{l,\lambda}\Big)^{n+m}
\int_{0}^{2\pi}\frac{d\theta}{2\pi}\mathcal{N}_{l,\lambda}(r^2)e^{i(n-m)
\theta}|re^{i\theta}\rangle_{l,\lambda}\,_{l,\lambda} \langle re^{i\theta}|\Bigg\}_{r=0}.
\end{eqnarray}
Therefore,  we can rewrite the density matrix (\ref{bel}) in  the form
\begin{eqnarray}
\label{esbel}
\rho&=&\sum_{n,m=0}^\infty\rho(n,m) 
\Bigg(\frac{q^{({}^{n+m}_{\;\;\;2})+nm}}{ \gamma^{n+m}(q^{1+n};q)_m(q^{1+m};q)_n}\Bigg)^{\frac{1}{2}}\cr
&\times&\Bigg\{\Big({_q}\partial_r^{l,\lambda}\Big)^{n+m}\int_{0}^{2\pi}\frac{d\theta}{2\pi}\mathcal{N}_{l,\lambda}(r^2)e^{i(m-n)\theta}
|re^{i\theta}\rangle_{l,\lambda}\,_{l,\lambda} \langle re^{i\theta}|\Bigg\}_{r=0},
\end{eqnarray}
generalizing the  Sudarshan's diagonal representation of the density matrix \cite{sudar} recovered in this case when $q\to1$, $l=1$ and $\lambda= 0.$

As stated in  \cite{sudar}, the form (\ref{esbel}) is particularly interesting since if $\mathcal{O}= (a^\dag)^n(a)^m$
is any normal ordered operator, its expectation value in the statistical state represented by the density matrix in the diagonal form expressed in terms of 
$(q;l,\lambda)-$coherent states 
\begin{eqnarray}
\label{fran}
\rho=\int d^2 z \phi(z) |z\rangle_{l,\lambda}\,_{l,\lambda} \langle z|,
\end{eqnarray}
is provided by
\begin{eqnarray}
tr(\rho\mathcal{O} )= tr(\rho(a^\dag)^n(a)^m)= \int_{
\mathbf{D}_{l,\lambda}}d^2z \phi(z) \bar{z}^n z^m,
\end{eqnarray}
where $d^2z= dRe(z)dIm(z)$ and $\int d^2 z \phi(z):=1$ (to ensure that $tr\rho=1$).  
The expansion 
coefficients  (\ref{bel}) and  (\ref{fran}) are expressed by the formula
\begin{eqnarray}
\label{franddc}
\quad 
\rho(n,m)=\frac{q^{\frac{1}{2}({}^n_2)+
\frac{1}{2}({}^m_2)}}{\sqrt{\gamma^{n+m}(q;q)_n(q;q)_m}}
\int \frac{d^2z\, \phi(z)}{\mathcal{N}_{l,\lambda}(|z|^2)} z^n\bar{z}^m,
\end{eqnarray}
which is the $(q;l,\lambda)-$analogue of  eq. (7) 
in \cite{sudar}. 
In polar coordinates $z=re^{i\theta},$ these coefficients
take the form
\begin{eqnarray}
\label{frandc}
\rho(n,m)=\frac{\pi\,q^{({}^n_2)}\,\delta_{n,m}}{\gamma^{n}(q;q)_n}
\int_{
0}^\infty \frac{d r^2\, \phi(r^2)}{\mathcal{N}_{l,\lambda}(r^2)} r^{2n}, \quad \mbox{ if } \quad 0 < q <1,
\end{eqnarray}
 and 
\begin{eqnarray}
\label{secon}
\rho(n,m)&=&\frac{\pi\, q^{({}^n_2)}\;\delta_{n,m}}{\gamma^{n}(q;q)_n}
\int_0^{R_{l,\lambda}} \frac{{_q}d_x^{l,\lambda}\,  \phi(x)}
{\mathcal{N}_{l,\lambda}(x) }x^{n},  \quad \mbox{ if } \quad  q >1.
\end{eqnarray}
The relation (\ref{secon}) generalizes the formula given by  Parthasarathy and Sridhar \cite{part1} which is recovered in our case by setting   $l= 1$ and $\lambda= 1.$
It follows from  \eref{frandc} or \eref{secon} that
\begin{eqnarray}
\label{frc}
\sum_{n=0}^\infty\rho(n,n)=1.
\end{eqnarray}

\subsection{Reproducing kernel and related properties}
 Consider  the  matrix  elements  of $\rho$ in the
deformed CS
\begin{eqnarray}
\label{arrr}
_{l,\lambda}\langle z'|\rho|z\rangle_{l,\lambda}=\sum_{n,m=0}^\infty\rho(n,m)\frac{q^{\frac{1}{2}({}^n_{2})+\frac{1}{2}({}^m_{\;2})}\,\bar{z}^{' n}z^m}{\sqrt{\gamma^{n+m}(q;q)_n(q;q)_m\mathcal{N}_{l,\lambda}(|z|^2)\mathcal{N}_{l,\lambda}(|z'|^2)}}.
\end{eqnarray}
\begin{definition}
Let $\mathbf{D}_{l,\lambda}$ be the open disc in $\mathbb{C}$ considered  in (\ref{radius}). Then, define the function
\begin{eqnarray}
\label{tat}
\rho: \mathbf{D}_{l,\lambda}\times\mathbf{D}_{l,\lambda}\longmapsto\mathbb{C},\quad (z',z)\longmapsto \rho(z',z),
\end{eqnarray}
\begin{eqnarray}
\label{azs}
\rho(z',z)&:=&\sum_{n,m=0}^\infty\rho(n,m)\frac{q^{\frac{1}{2}({}^n_2)+\frac{1}{2}({}^m_{\;2})}}{\sqrt{\gamma^{n+m}(q;q)_n(q;q)_m}}\frac{\bar{z}^{' n}z^m}{\sqrt{\mathcal{N}_{l,\lambda}(|z|^2)\mathcal{N}_{l,\lambda}(|z'|^2)}}.
\end{eqnarray}
\end{definition}
\begin{proposition}
The function $\rho$ (\ref{azs}) can be re-expressed as follows
\begin{eqnarray}
\label{tate}
\rho(z',z)=\int_{
\mathbf{D}_{l,\lambda}} d^2\zeta K(\zeta,z)\rho(z',\zeta),
\end{eqnarray}
where
\begin{eqnarray}
\label{sasa}
K(z,\zeta)
=\frac{_{l,\lambda}\langle \zeta|z\rangle_{l,\lambda}}{ \pi\ln q^{-1}}\frac{1}{\eta+|\zeta|^2}, \quad \mbox{ if } \quad 0 < q <1,
\end{eqnarray}
and 
\begin{eqnarray}
\label{sasasa}
K(z,\zeta)
=\frac{_{l,\lambda}\langle \zeta|z\rangle_{l,\lambda}}{ \pi(1+\frac{x}{\eta})},
\quad \mbox{ if }\quad  q>1,\;0 < x=|\zeta|^2 < \frac{l^2q^\lambda}{q-1},\;\theta=\arg\zeta.
\end{eqnarray}
\end{proposition}
{\bf Proof.} Using  (\ref{resolv}), the relation (\ref{azs}) is readily put in the form
\begin{eqnarray}
\label{tatesq}
\rho(z',z):={_{l,\lambda}}\langle z'|\rho|z\rangle_{l,\lambda}=
 \int_{
\mathbf{D}_{l,\lambda}} d^2\zeta \rho(z',\zeta)K(\zeta,z),
\end{eqnarray}
where
$
K(z,\zeta)
=\frac{_{l,\lambda}\langle \zeta|z\rangle_{l,\lambda}}{ \pi\ln q^{-1}}\frac{1}{\eta+|\zeta|^2} \; \mbox{ if }\; 0 < q <1
$
and 
$
K(z,\zeta)
=\frac{_{l,\lambda}\langle \zeta|z\rangle_{l,\lambda}}
{ \pi(1+\frac{x}{\eta})}\; \mbox{ if }\;  q>1,\; 0 < x=|\zeta|^2 < \frac{l^2q^\lambda}{q-1},\;\theta=arg(\zeta).
$
$\square$

The equation (\ref{tate}) encodes nothing but  the reproducing property 
of $\rho(z',z) $  where $K(z,\zeta)$ is the reproducing kernel. Indeed, we get:
\begin{proposition}
The  quantity $K(z,\zeta)$ given in (\ref{sasa}) and  (\ref{sasasa})
\begin{enumerate}
\item satisfies the matrix multiplication property, i.e, 
\begin{eqnarray}
\label{tte}
\int_{
\mathbf{D}_{l,\lambda}} d^2\zeta K(z,\zeta)K(\zeta,z')=K(z,z'),
\end{eqnarray}
\item obeys the Hermiticity property, i.e, 
$
(K(z,\zeta))^*=K(\zeta,z),
$
\item is positive, 
\item is  an entire function, i.e, for  $t=(q-1)|z|^2/l^2q^\lambda$
\begin{eqnarray}
K(z,z)=\frac{1-q}{l^2q^\lambda\pi\ln q^{-1}}
\sum_{n=0}^\infty t^n,\quad \mbox{ if } \quad 0 < q <1,
\end{eqnarray}
and 
\begin{eqnarray}
K(z,z)=\frac{1}{\pi}
\sum_{n=0}^\infty t^n, \quad \mbox{ if } \quad q>1.
\end{eqnarray}
\end{enumerate}
\end{proposition}
{\bf Proof.} 

$\bullet$
From the resolution of the  identity (\ref{resolv}), we have
\begin{eqnarray*}
\int_{
\mathbf{D}_{l,\lambda}} d^2\zeta K(z,\zeta)K(\zeta,z')
=\frac{1 }{ \pi\ln q^{-1}}\frac{_{l,\lambda}\langle z'|z\rangle_{l,\lambda}}{\eta+|z|^2}\int_{
\mathbf{D}_{l,\lambda}}
\frac{d^2\zeta }{ \pi\ln q^{-1}}\frac{|\zeta\rangle_{l,\lambda}\,_{l,\lambda}\langle \zeta|}{\eta+|\zeta|^2}
\end{eqnarray*}
if $0< q < 1$ and 
 \begin{eqnarray*}
\int_{
\mathbf{D}_{l,\lambda}} d^2\zeta K(z,\zeta)K(\zeta,z')=\frac{1 }{ \pi}\frac{_{l,\lambda}\langle z'|z\rangle_{l,\lambda}}{1+r/\eta}\int_{
\mathbf{D}_{l,\lambda}}
\frac{d^2\zeta }{ \pi}\frac{|\zeta\rangle_{l,\lambda}\,_{l,\lambda}\langle \zeta|}{1+x/\eta}
\end{eqnarray*}
if $q>1.$ By setting $|\zeta|^2=x,\;|z|^2=r,\; \eta=l^2q^\lambda/(1-q)$ and  using the overcompleteness of the $(q;l,\lambda)-$coherent states, the proof of (i) is achieved.

$\bullet$ (ii),  (iii) and (iv) ($z=\zeta$) are immediate from   equations (\ref{sasa}) and (\ref{sasasa}).
 $\square$

In the limit $q\rightarrow 1$ when  $l=1,$  the expression (\ref{sasasa})
is reduced to the undeformed reproducing kernel  given in \cite{acc}. Further, the reproducing kernel  
possesses  required properties as stated in the following.

Let now $F$ and  $\tilde{F}$ be the normal ordering and   anti-normal ordering  operators,   defined as $F:=a^{\dag \sigma}a^\nu$ and $ \tilde{F}:=a^\nu a^{\dag \sigma}, $ respectively. 
Then, as treated above,  the  expectation value of $F$  is  given by
\begin{eqnarray}
\label{expeddc}
tr(\rho F)=\int_{
\mathbf{D}_{l,\lambda}}d^2z \phi(z) \bar{z}^\sigma z^\nu.
\end{eqnarray}
This  can be identified as the  expectation value of the complex classical function $z^\nu\bar{z}^\sigma$
for a probability distribution $\phi(z)$
over the complex domain $\mathbf{D}_{l,\lambda}.$ For $\phi(z)=\phi_1(\theta)\phi_2(r),$
\eref{expeddc} takes the form
\begin{eqnarray}
\label{expece}
tr(\rho F)=\int_0^{2\pi}\frac{d\theta}{2} e^{i(\nu-\sigma)}\phi_1(\theta)
\int_0^\infty
d r^2 \phi_2(r) r^{2\nu} \quad \mbox{ if }\quad  0< q< 1,
\end{eqnarray}
and
\begin{eqnarray}
\label{expece}
tr(\rho F)=\int_0^{2\pi}\frac{d\theta}{2} e^{i(\nu-\sigma)}\phi_1(\theta)
\int_0^\infty
{_q}d_x^{l,\lambda}  \phi_2(x) x^{\nu} \quad \mbox{ if }\quad   q> 1.
\end{eqnarray}
As matter of concrete illustration, consider now  a $(q;l,\lambda)-$deformed Gaussian function $\phi_2(r)=\frac{1}{\pi}\mathcal{N}_{l,\lambda}^{-1}(r^2)$ with $\phi_1(\theta)=1$ and $\phi_2(x)=\frac{1}{\pi}\mathcal{N}_{l,\lambda}^{-1}(x)$ with  $\phi_1(\theta)=1$, in the above equations, respectively. Then we get
\begin{eqnarray}
\label{trace}
tr(\rho F)=\ln q^{-1}\gamma^{\nu+1} q^{-({}^{\nu+1}_{\;\;2})}(q;q)_\nu,\quad \mbox{ if }\quad  0< q< 1.
\end{eqnarray}
and 
\begin{eqnarray}
\label{trace1}
tr(\rho F)=(-\gamma)^{\nu} q^{-1}(q^{-1};q^{-1})_\nu,\quad \mbox{ if }\quad q>1.
\end{eqnarray}

Similarly to \eref{expeddc},  the  expectation value of $\tilde{F}$ is  given by
\begin{eqnarray}
\label{expec}
tr(\rho \tilde{F})=\int_{
\mathbf{D}_{l,\lambda}}d^2z \phi(z) \,_{l,\lambda}\langle z|a^\nu a^{\dag \sigma}|z\rangle_{l,\lambda}.
\end{eqnarray}
Turning back to  polar coordinates with $z=\sqrt x e^{i\theta}$, and assuming  $\phi(z)=\phi_1(\theta)\phi_2(x)$. 
the latter expression is explicitly evaluated as
\begin{eqnarray}
\label{exc}
\fl
tr(\rho \tilde{F})=q^{-({}^\nu_2)}\gamma^\nu(q;q)_\nu\sum_{n=0}^\infty
\frac{q^{({}^n_2)-n\nu}(q^{1+\nu};q)_n}{(q,q;q)_n\gamma^n}\int_0^{2\pi}\frac{d\theta}{2}
e^{i(\nu-\sigma)}\phi_1(\theta)\int_0^{R_{l,\lambda}} x^n
\frac{\phi_2(x){_q}d_x^{l,\lambda} }{\mathcal{N}_{l,\lambda}(x)}. 
\end{eqnarray}
 For now $\phi_1(\theta)=1,\;\phi_2(x)=1/\pi\mathcal{N}_{l,\lambda}(x),$
\begin{eqnarray}
\label{expeesc}
tr(\rho \tilde{F})=q^{-({}^\nu_2)-1}\gamma^\nu
(q;q)_\nu(q^{-1};q^{-1})_\infty^2
\mathcal{O}_\infty(-q^{-\nu};q^{1+\nu}|q),\;\;q>1,
\end{eqnarray}
with
\begin{eqnarray}
 \mathcal{O}_\infty(x;q^{1+m}|q):=\sum_{n=0}^\infty \frac{q^{({}^n_2)}(q^{1+m};q)_n}{(q,q;q)_n}x^n
J_0^{(1)}(2iq^{-(1+n)/2};q^{-1}),
\end{eqnarray}
where the function $J_0^{(1)}(z;q):= 
\sum_{n=0}^{\infty}\frac{(-1)^n(\frac{z}{2})^{2n}}{(q,q;q)_n}$ is the $q-$deformed Bessel function \cite{ASK}.
 
Provided the above results, it becomes easier to compute the expectation value of the  deformed harmonic oscillator  Hamiltonian operator $H$ in the following interesting cases: 
\begin{enumerate}
\item  $H=a^\dag a:$ Exploiting  \eref{trace} and \eref{trace1} leads to
\begin{eqnarray}
tr(\rho H)=\gamma^2 q^{-1}(1-q)\ln q^{-1},\quad \mbox{ if }\quad  0< q< 1,
\end{eqnarray}
and 
\begin{eqnarray}
tr(\rho H)=l^2q^{\lambda-3},\quad \mbox{ if }\quad q>1.
\end{eqnarray}
\item $H=a a^\dag+a^\dag a: $ Using \eref{trace}, \eref{trace1} and \eref{exc} yields:
\begin{eqnarray}
 tr(\rho H)=\left\{\begin{array}{ll}l^2q^{\lambda-1}\Big(1+\ln q^{-2}\frac{l^2q^{\lambda-2}}{1-q}\Big),& \mbox{ if } \;0< q< 1,\\
 l^2q^{\lambda-1}(1+2q^{-2}),&\mbox{ if } \,  q> 1.
\end{array}
\right.
\end{eqnarray}
\end{enumerate}
In the same vein,  the  expectation values of the deformed position and momentum operators translate into
\begin{eqnarray}
\label{exxx}
tr(\rho Q)=\sqrt 2\int_{
\mathbf{D}_{l,\lambda}}d^2z \phi(z) Re(z),\; tr(\rho P)=\sqrt 2\int_{
\mathbf{D}_{l,\lambda}}d^2z \phi(z) Im(z).
\end{eqnarray}
Suppose for simplicity  $\phi(z)=\phi_{1,i}(\theta)\phi_{2,i}(|z|),\;\varphi_i(r,\theta)$ and $\tilde{\varphi}_i(x,\theta)$ such that
\begin{eqnarray}
\varphi_i(r,\cos\theta):=\int_0^{2\pi}\frac{d\theta}{\sqrt 2}\cos\theta\,\phi_{1,i}(\theta)\int_0^\infty rd r^2 \phi_{2,i}(r),\;0 < q <1,\\
\tilde{\varphi}_i(x,\cos\theta):=\int_0^{2\pi}\frac{d\theta}{\sqrt 2}\cos\theta\,\phi_{1,i}(\theta)\int_0^{R_{l,\lambda}}x^{1/2}\phi_{2,i}(x){_qd}_x^{l,\lambda},\;q>1,
\end{eqnarray}
where $i=Q,\;P$
and use the polar coordinates   to detail expressions in \eref{exxx} as follows:
\begin{eqnarray}
\label{exx}
tr(\rho Q)=\varphi_Q(r,\cos\theta),\qquad 
 tr(\rho P)=\varphi_P(r,\sin\theta)
\end{eqnarray}
if $0< q<1,\; |z|=r,\;0\leq\theta\leq2\pi,\;d^2z=dRe(z) dIm(z)$
and
\begin{eqnarray}
\label{exerx}
tr(\rho Q)=\tilde{\varphi}_Q(x,\cos\theta),\qquad
 tr(\rho P)=\tilde{\varphi}_P(x,\sin\theta)
\end{eqnarray}
if $q>1,\; 0< |z|^2=x<R_{l,\lambda},\;0\leq\theta\leq2\pi,d^2z=\frac{1}{2}{_qd}_x^{l,\lambda}d\theta.$ Therefore,
\begin{enumerate}
\item For the simpler case of $\phi_{1,Q}(\theta)=\phi_{1,P}(\theta)= 1$, the  expectation values of $Q$ and $P$ are reduced to zero, i.e 
$$tr(\rho Q)=tr(\rho P)=0.$$
\item For $\phi_{1,Q}(\theta)=1/\sqrt{2}\cos\theta,\;\phi_{1,P}(\theta)=1/\sqrt{2}\sin\theta$, the  expectation values of $Q$ and $P$ 
  can be re-expressed in terms of $\phi_{2,i}(|z|),$ i.e
\begin{eqnarray}
\fl
\qquad tr(\rho Q)=\pi\int_0^\infty rd r^2 \phi_{2,Q}(r),\quad tr(\rho P)=\pi\int_0^\infty rd r^2 \phi_{2,P}(r)
\end{eqnarray}
if $0 < q <1$ and
\begin{eqnarray}
\label{lan}
tr(\rho Q)=\pi\Big(\frac{l^2q^\lambda}{q-1}\Big)^{1/2}\sum_{n=0}^\infty
q^{-3n/2}\phi_{2,Q}\Big(\frac{l^2q^{\lambda-n}}{q-1}\Big),\\
tr(\rho P)=\pi\Big(\frac{l^2q^\lambda}{q-1}\Big)^{1/2}\sum_{n=0}^\infty
q^{-3n/2}\phi_{2,P}\Big(\frac{l^2q^{\lambda-n}}{q-1}\Big)
\end{eqnarray}
if $q>1.$

 Assigning now concrete expressions to the unknown functions, e.g. $\phi_{2,Q}(r)=\phi_{2,P}(r)\equiv 1/\pi\mathcal{N}_{l,\lambda}(r)$,  the  expectation values of the deformed position and momentum operators coincide and give 
\begin{eqnarray}
tr(\rho Q)=tr(\rho P)=\left\{\begin{array}{ll}\Big(\frac{l^2q^{\lambda-2}}{1-q}\Big)^3\ln q^{-1}(q;q)_2,&\; 0< q<1,\\\\
\Big(\frac{l^2q^{\lambda-3}}{q-1}\Big)^{1/2}\frac{(q^{-1};q^{-1})_\infty}{(q^{-3/2};q^{-1})_\infty}, &\; q>1.
\end{array}
\right.
\end{eqnarray}
\end{enumerate}

To end this discussion, let us compute  the  matrix elements and expectation value for the Hamiltonian operator describing the propagation of light in a non-linear medium
like Kerr medium.  Such a Hamiltonian is usually expressed by \cite{sudar}
\begin{eqnarray}
\label{actt}
H_d:=a^\dag a+\frac{\chi}{2}a^{\dag 2} a^2,
\end{eqnarray}
by setting  $\hbar=1=\omega; $ $\chi$ represents the interaction strength of the light with the non-linear medium. Its  matrix elements in the Fock space states are provided by
\begin{eqnarray}\label{kerr}
\langle r|H_d|s\rangle=\varphi(s)\Big(1+\frac{\chi}{2 }\varphi(s-1)\Big)\delta_{r,s},
\end{eqnarray}
while its expectation value is evaluated as follows:
\begin{eqnarray}
 tr(\rho H_d)=\left\{\begin{array}{ll}\gamma^2q^{-1}(1-q)\ln q^{-1}\Big\{1+\frac{\chi}{2}\gamma q^{-2}(1-q^2)\Big\},& \mbox{ if } \;0< q< 1\\
l^2q^{\lambda-3}\{1+\frac{\chi}{2}l^2q^{\lambda-3}(1+q)\},&\mbox{ if } \,  q> 1.\end{array}
\right.
\end{eqnarray}
We observe that the matrix elements (\ref{kerr}) strongly depends on the algebra structure function, and hence on the deformed number operator. A non-negligible Kerr effect can be exhibited in such a non-linear medium  depending on how much is the contribution of the interaction strength of the light with it.  
\section{Relevant matrix elements and  mean values}
This section deals with the computation of relevant   normal and anti-normal forms in the coherent states $|z\rangle_{l,\lambda}$ and their relation with engendered new deformed $(q; l, \lambda)-$hypergeometric functions.  As a first step in such a direction, the following preliminary result stated as a lemma  reveals to be useful.

\begin{lemma}
\label{ma}
The matrix elements of the normal  form 
are given by
\begin{eqnarray}
\label{eees}
\fl
\qquad\;\;\langle r|a^{\dag m}a^n|s\rangle=
(- q)^n(q^{-s};q)_nq^{-\frac{1}{2}({}^{m-n}_{\;\;\;2})-\frac{s(m-n)}{2}}
\sqrt{\gamma^{m+n}(q^{1+s};q)_{m-n}}\;\delta_{r,m-n+s},
\end{eqnarray}
if $ n< m$,
\begin{eqnarray}
\label{scend}
\fl
\qquad\langle r|a^{\dag m}a^n|s\rangle=
(- q)^m(q^{-s-m+n};q)_mq^{-\frac{1}{2}({}^{n-m}_{\;\;\;2})-\frac{s(n-m)}{2}}
\sqrt{\gamma^{m+n}(q^{1+r};q)_{n-m}}\;\delta_{s,n-m+r},\cr
\end{eqnarray}
if $n>m$,\\
  while those of  the  anti-normal form are expressed by
\begin{eqnarray}
\label{eele}
\langle r|a^{n}a^{\dag m}|s\rangle=
q^{-\frac{1}{2}({}^{n}_2)-\frac{1}{2}({}^{m}_{\,2})-\frac{rn+sm}{2}}
\sqrt{\gamma^{n+m}(q^{1+r};q)_{n}(q^{1+s};q)_{m}}\;\delta_{r+n,s+m}.
\end{eqnarray}
\end{lemma}
{\bf Proof.}  See appendix B. 
\begin{proposition}
\label{pppp}
The expectation values of the normal and anti-normal forms  in the coherent states $|z\rangle_{l,\lambda}$ are given, respectively, by 
\begin{eqnarray}
\label{matrixe}
\langle a^n a^{\dag m}\rangle
=(q^{-m};q)_n\frac{(-\gamma q)^n\bar{z}^{m-n}}{\Big(-\frac{|z|^2}{\gamma};q\Big)_\infty}\;{_1}\phi_1\left(
\begin{array}{c}q^{1+m}\\
q^{1+m-n}\end{array}\Bigg|q;-\frac{|z|^2q^{-n}}{\gamma}\right)
\end{eqnarray}
if $n<m,$
\begin{eqnarray}
\label{mtrxe}
\langle a^na^{\dag m}\rangle
=
(q^{-n};q)_m\frac{(-\gamma q)^m z^{n-m}}{\Big(-\frac{|z|^2}{\gamma};q\Big)_\infty}\;{_1}\phi_1\left(
\begin{array}{c}q^{1+n}\\
q^{1+n-m}\end{array}\Bigg|q;-\frac{|z|^2q^{-m}}{\gamma}\right)
\end{eqnarray}
if $n>m,$ and
\begin{eqnarray} 
\label{mtrxee}
\langle a^n a^{\dag n}\rangle
=
q^{-({}^n_2)}\gamma^n(q;q)_n\Big(-\frac{|z|^2}{\gamma};q\Big)_\infty^{-1}\;{_1}\phi_1\left(
\begin{array}{c}q^{1+n}\\
q\end{array}\Bigg|q;-\frac{|z|^2q^{-n}}{\gamma}\right).
\end{eqnarray}
 Moreover, for any two integers $n$ and $m,$
\begin{eqnarray}
\label{msazz}
\langle a^{\dag m}a^n\rangle
=\bar{z}^mz^n,
\end{eqnarray}
where 
\begin{eqnarray}
 {_1}\phi_1\left(\begin{array}{c}
A\\B\end{array}\Bigg|q;t\right)=\sum_{n=0}^\infty\frac{(-1)^nq^{({}^n_2)}(A;q)_n}{(B,q;q)_n}t^n
\end{eqnarray}
 is the $q-$deformed hypergeometric function \cite{ASK}
 and $\gamma=l^2q^{\lambda-1}/(1-q)$.
\end{proposition}
{\bf Proof.} By definition 
\begin {eqnarray}
\label{rp}
\fl
\qquad\langle a^na^{\dag m}\rangle
:&=&\frac{_{l,\lambda}\langle z| a^na^{\dag m}|z\rangle_{l,\lambda}}{_{l,\lambda}\langle z|z\rangle_{l,\lambda}}\cr
&=&\mathcal{N}_{l,\lambda}^{-1}(|z|^2)\sum_{r,s=0}^\infty\frac{q^{({}^r_2)+({}^s_2)}\bar{z}^rz^s}{\gamma^{r+s}(q;q)_r(q;q)_s}\langle 0|a^{r+n}a^{\dag m+s}|0\rangle.
\end{eqnarray}
By using (\ref{ket}), the equation (\ref{rp}) can be rewritten as
\begin {eqnarray}
\label{rp1}
\fl
\qquad\langle a^na^{\dag m}\rangle
&=&\mathcal{N}_{l,\lambda}^{-1}(|z|^2)\sum_{r,s=0}^\infty\frac{q^{({}^r_2)+({}^s_2)}\bar{z}^rz^s}{\gamma^{r+s}(q;q)_r(q;q)_s}q^{-\frac{1}{2}({}^{r+n}_{\;\;\;2})-\frac{1}{2}({}^{s+m}_{\;\;\;2})}\cr
&\times&\sqrt{\gamma^{r+n+s+m}(q;q)_{r+n}(q;q)_{s+m}}\;\delta_{r+n,s+m}\cr
&=&\mathcal{N}_{l,\lambda}^{-1}(|z|^2)(-\gamma q)^n \bar{z}^{m-n}(q^{-m};q)_n\sum_{s=0}^\infty\frac{q^{({}^s_2)} (q^{1+m};q)_s}{(q^{1+m-n},q;q)_s}\Bigg(\frac{|z|^2}{\gamma}q^{-n}\Bigg)^s.
\end{eqnarray}
The proof of (\ref{matrixe}) is achieved by re-expressing the sum in terms of a formula analog to $q-$hypergeometric function.  
 Idem for the other expressions. $\square$

\begin{proposition}
From the above derived results,  follow the properties:
\begin{itemize}
\item[(i)] 
\begin{eqnarray}
\langle a \rangle=z,\quad \langle  a^{\dag  }\rangle=\bar{z},\quad\langle Q \rangle=\sqrt2\,Re(z),\quad \langle  P\rangle=\sqrt2\,Im(z),
\end{eqnarray}
\item[(ii)] 
\begin{eqnarray}
\fl
\quad \langle \varphi(N) \rangle:=\langle a^\dag a \rangle=|z|^2,\; \langle  a a^{\dag  }\rangle=l^2q^{\lambda-1}+q^{-1}|z|^2,\; \langle   H\rangle=l^2q^{\lambda-1}+(1+q^{-1})|z|^2,\cr
\end{eqnarray}
\item[(iii)]
\begin{eqnarray}
\label{samma}
 \langle Q^2\rangle=\frac{3+q^{-1}}{2}Re^2(z)+\frac{q^{-1}-1}{2}Im^2(z)+\frac{l^2q^{\lambda-1}}{2},\\
\label{sammaa}\langle P^2\rangle=\frac{q^{-1}-1}{2}Re^2(z)+\frac{3+q^{-1}}{2}Im^2(z)+\frac{l^2q^{\lambda-1}}{2}.
\end{eqnarray}
\item[(iv)]
\begin{eqnarray}
 (\Delta Q)^2=(\Delta P)^2=\Delta Q \Delta P=\frac{l^2q^{\lambda-1}}{2}+\frac{q^{-1}-1}{2}|z|^2.
\end{eqnarray}
\end{itemize}
\end{proposition}
{\bf Proof.} The proof is immediate by using (\ref{matrixe}), (\ref{mtrxe}) and (\ref{mtrxee}). $\square$

As we can see from these relations, the mean values of the ladder operators $a$ and $a^\dag$ and the deformed number operator $\varphi(N)$ conserve  the same values $z$, $\bar{z}$ and $|z|^2,$ respectively, as in standard coherent states \cite{gazeau09}.  In opposite, the mean values of
$ \langle  a a^{\dag  }\rangle$ and $H$ are strongly affected by the algebra deformation parameters $q, l, \lambda$. Moreover, although a coherent state is an infinite superposition of Fock states $|n\rangle$, one
gets a finite mean value of $N$ that can be arbitrarily small at $z\rightarrow 0.$
\section{Berezin-Klauder-Toeplitz  quantization}
The Berezin-Klauder-Toeplitz quantization, (also called ''anti-Wick" or coherent state quantization), of  phase space observables of the  complex plane, $\mathbf{D}_{l,\lambda},$  uses the resolution of the identity \eref{resolv} and  is performed by 
mapping a function $f$ that satisfies appropriate conditions, to  the following operator in the Hilbert space  (\cite{klauder63, gazeautt} and references therein):
\begin{eqnarray}
\label{tota1}
 f\longmapsto A_f=\int_{\mathbf{D}_{l,\lambda}}d\mu_{l,\lambda}(\bar{z},z)f(z,\bar{z})|z\rangle_{l,\lambda}\,_{l,\lambda}\langle z|=\sum_{n,n'=0}^\infty\left(A_F\right)_{nn'}|n\rangle\langle n'|,
\end{eqnarray}
where this integral is understood in the weak sense, i.e. it defines in fact a sesquilinear form (eventually only densely defined) 
\begin{eqnarray*}
B_f(\psi_1,\psi_2)= \int_{\mathbf{D}_{l,\lambda}}d\mu_{l,\lambda}(\bar{z},z) f(z,\bar{z})
\,\langle \psi_1|z\rangle_{l,\lambda}\,_{l,\lambda}\langle z|\psi_2\rangle,
\end{eqnarray*}
with the  matrix elements 
\begin{eqnarray}
\label{matr}
\left(A_F\right)_{nn'}=\frac{q^{n(n-1)/4+n'(n'-1)/4}}{\sqrt{\gamma^{n+n'}(q;q)_n(q;q)_{n'}}}\int_{\mathbf{D}_{l,\lambda}}f(z,\bar{z})z^{n}\bar{z}^{n'}\frac{d\mu_{l,\lambda}(\bar{z},z)}{\mathcal{N}_{l,\lambda}(|z|^2)}.
\end{eqnarray}
Operator $A_f$ is symmetric if $f(z,\bar{z})$ is real-valued,  and is bounded (resp. semi-bounded)
if $f(z,\bar{z})$  is bounded (resp. semi-bounded). In particular, the Friedrich extension allows
to define $A_f$ as a self-adjoint operator if $f(z,\bar{z})$ is a semi-bounded real-valued function.
Note that the original $f(z,\bar{z})$ is a “upper or contravariant symbol”, usually non-unique, for the operator
$A_f.$ 
This problem involving the property of the function $f$ and the self-adjointness criteria of operators is thoroughly discussed  in a recent work by Bergeron et al \cite{Bergeron} and does not deserve further development here.
So, without loss of generality, let us immediately examine different concrete expressions  for the function $f$ in the line of  \cite{gazeautt} as matter of result comparison:
\begin{enumerate}
 \item The function $f$  only depends on $|z|^2 =t:$  the matrix elements \eref{matr} take  the form 
\begin{eqnarray*}
\left(A_F\right)_{nn'}=\frac{q^{n(n+1)/2}(1-q)\,\delta_{n,n'}}{\ln q^{-1}[n]_q!}\int_{0}^\infty dt\; \frac{t^n\,f(tl^2q^\lambda)}{E_q((1-q)t)},\quad \mbox{ if } \; 0 < q <1,
\end{eqnarray*}
 and
\begin{eqnarray*}
\left(A_F\right)_{nn'}=\frac{q^{n(n+1)/2}\,\delta_{n,n'}}{[n]_q!}\int_{0}^{l^2q^\lambda/(q-1)} dt\; t^n\frac{f(tl^2q^\lambda)}{E_q((1-q)t)},\quad \mbox{ if } \;q>1.
\end{eqnarray*}
\item The function $f$ only  depends on the angle $\theta = \arg z,$  i.e. $f(z,\bar z) = F(\theta):$   the matrix elements \eref{matr}  are given by 
\begin{eqnarray*}
\label{matelang1}
\left(A_F\right)_{nn'}= c_{n'-n}(F) \Bigg(\frac{q^{({}^n_2)+({}^{n'}_{\,2})-\frac{(n+n')(n+n'-2)}{4}}}{(q;q)_n(q;q)_{n'}}\Bigg)^{1/2}\;(q;q)_{\frac{n+n'}{2}},
\end{eqnarray*}
where  $c_n(F)$ are the Fourier coefficients of the function $F:$   
$$
c_n(F) = \displaystyle \frac{1}{2\pi}\int_0^{2\pi}d\theta\, e^{-in\theta}\, F(\theta) ,
$$
while the self-adjoint ``angle'' operator is defined by
 \begin{eqnarray}
\label{angleop1}
A_{\theta}= \pi\,  I_{{\mathcal F}} + i \, \sum_{n\neq n'} \Bigg(\frac{q^{({}^n_2)+({}^{n'}_{\,2})-\frac{(n+n')(n+n'-2)}{4}}}{(q;q)_n(q;q)_{n'}}\Bigg)^{1/2} \frac{(q;q)_{\frac{n+n'}{2}}}{n'-n}\, |n\rangle\langle n'|.
\end{eqnarray}
\item The function $f(z,\bar{z})=z$ and $f(z,\bar{z})=\bar{z}:$ the operator
\eref{tota1} takes the forms
\begin{eqnarray}
\label{stoper1}
   A_z &=& a, \; \quad a \, | n \rangle = \sqrt{\varphi(n)} | n-1\rangle\, ,
   \quad\; a\,| 0\rangle  = 0,\\
 \label{stoper2}
    A_{\bar z} & = &a^{\dag}, \quad a^{\dag} \, | n \rangle =
     \sqrt{\varphi(n+1)} |n+1\rangle, 
\end{eqnarray}
where the function $\varphi$ is defined in \eref{rela}. The state $|z\rangle_{l,\lambda}$ is eigen-vector of $A_z = a$ with eigenvalue $z$ 
like for standard CS. 
The operators $A_z$ and $A_{\bar{z}}$ satisfy the algebra \eref{rela}, i.e
$$
[A_z,A_{\bar{z}}]=l^2q^{\lambda-1-N},
$$
as required.
\item  The function $f(z,\bar{z})=z^\mu\bar{z}^\nu,\;\mu, \nu\in\mathbb{N}\cup \{0\}:$ the matrix elements \eref{matr} of $A_f$ are given by
\begin{eqnarray}
\label{matelan}
\fl 
\left(A_F\right)_{nn'}= \Bigg(\frac{q^{({}^n_2)+({}^{n'}_{\,2})-\frac{(n+n'+\mu+\nu)(n+n'+\mu+\nu-2)}{4}}\gamma^{\mu+\nu}}{(q;q)_n(q;q)_{n'}}\Bigg)^{1/2}\;(q;q)_{\frac{n+n'+\mu+\nu}{2}}\;\delta_{n-n',\nu-\mu}.
\end{eqnarray} 
\item The function $f$ is defined on the complex plane $\mathbb{C}=\Big\{z=\frac{{\bf q} + i{\bf p}}{\sqrt 2}\Big\},$ i.e 
$f(z,\bar{z}) = \vert z \vert^2$. 
In this case, $f= \vert z \vert^2 = \frac{1}{2}({\bf p}^2 + {\bf q}^2)$ looks like  the classical Hamiltonian of the harmonic oscillator with $\omega=1=2m$ where $m$ is the particle mass. Moreover, 
\begin{equation}
\label{oustase}
A_{(p^2+q^2)/2}=A_{z\bar z} = A_z\, A_{\bar z}= a\, a^{\dag} =\varphi(N+1).
\end{equation}

Therefore, the spectrum of the quantized version of $f=\vert z \vert^2 $  is the 
 sequence $\left\{ \varphi(n)\right\}_{n\geq 1}$.
\end{enumerate}

Now denoting by  $A_{\bf q}$ and $A_{\bf p}$  the quantized $(q; l, \lambda)-$deformed
position and momentum operators corresponding to the phase space coordinates ${\bf q}$ and ${\bf p}$, respectively, we have
\begin{equation}\label{QP-operators}
 A_{\bf q}:=Q = \frac 1{\sqrt{2}}(a^\dag+a), \qquad
 A_{\bf p}:= P = \frac{i}{\sqrt{2}}( a^\dag-a).
\end{equation}
and, in the case of quadratic expressions,
 \begin{equation}
\label{q2p2}
\fl
\qquad\qquad
A_{\bf q^2}= Q^2 + \frac{1}{2}(\varphi(N+1)-\varphi(N)), \;\; A_{\bf p^2}= P^2 + \frac{1}{2}(\varphi(N+1)-\varphi(N)),  
\end{equation}
 while the deformed harmonic oscillator Hamiltonian quantizes as follows: 
  \begin{equation}
\label{q2+p2}
\fl
\qquad \quad
\frac{1}{2}(P^2 + Q^2) = A_{({\bf p^2}+{\bf q^2})/2} -\frac{1}{2}(\varphi(N+1)-\varphi(N)) = \frac{1}{2}(\varphi(N+1)+\varphi(N)).   
\end{equation}

 The time evolution of the  quantized version $a=A_z$ of the classical phase space point $z=(  {\bf q} + i{\bf  p} )/\sqrt 2$  is given by
\begin{eqnarray}
\label{eev}
  \check{z}(t) :&=&{_{l,\lambda}}\langle
z| e^{-i\hat{H}t}\,  A_z \, e^{i\hat{H}t}|z\rangle_{l,\lambda}\cr
 &=&\frac{z}{\mathcal{N}_{l,\lambda}(|z|^{2})}\sum_{n=0}^{+\infty} 
\frac{q^{({}^n_2)}| z|^{2n}}{(q;q)_n\gamma^n}\,\exp(it l^2q^{\lambda-2-n}(1+q)).
\end{eqnarray}
In the limit when $q\to 1,$  $l=1/\sqrt 2,$ one recovers the standard case  $ \check{z}(t) =z e^{i t}$ which describes the classical phase-space trajectory.
For a fixed normalized state $|z_0\rangle_{l,\lambda}$ on $\mathbf{D}_{l,\lambda}\subset\mathbb{C}$, one can define the probability
distribution on the complex plane  by the map
\begin{eqnarray}
\label{probdensphs}
\fl
\quad 
\mathbb{C}\ni z=(q+ip)/\sqrt{2}\mapsto \rho_{|z_0\rangle_{l,\lambda}}(z) := \vert_{l,\lambda} \langle z|z_0\rangle_{l,\lambda} \vert^2= \frac{\vert \mathcal{N}_{l,\lambda}(\bar z z_0 )\vert^2}{\mathcal{N}_{l,\lambda}(\vert z\vert^2)\mathcal{N}_{l,\lambda}(\vert z_0\vert^2)}.
\end{eqnarray}
So, the 
time evolution behavior $t \mapsto\rho_{|z_0\rangle_{l,\lambda}}(z,t)$
 of the probability density  \eref{probdensphs} 
is given by
\begin{equation}
\label{probdens}
z \mapsto \rho_{|z_0\rangle_{l,\lambda}}(z,t):=\vert _{l,\lambda}\langle z| e^{-i\hat Ht}|z_0\rangle_{l,\lambda} \vert^2=\frac{\vert \mathcal{N}_{l,\lambda;t,\varphi}(\bar z z_0 )\vert^2}{\mathcal{N}_{l,\lambda}(\vert z\vert^2)\mathcal{N}_{l,\lambda}(\vert z_0\vert^2)},
\end{equation}
where the series $\mathcal{N}_{l,\lambda;t,\varphi}$ is defined as follows
\begin{eqnarray}
\mathcal{N}_{l,\lambda;t,\varphi}(x):=\sum_{n=0}^\infty\frac{q^{({}^n_2)}}{(q;q)_ne^{it\varphi(n+1)}}\Big(\frac{x}{\gamma}\Big)^n.
\end{eqnarray}
Defining the function
\begin{eqnarray}
\label{tat}
\check{f}: \mathbf{D}_{l,\lambda}\times\mathbf{D}_{l,\lambda}\longmapsto\mathbb{C},\quad (z,\bar{z})\longmapsto \check{f}(z,\bar{z}),
\end{eqnarray}
\begin{equation}
\label{wksssymb}
\check{f}(z,\bar{z}):={ _{l,\lambda}}\langle z| A_f |z\rangle _{l,\lambda}= \int_{\mathbf{D}_{l,\lambda}}d\mu_{l,\lambda}(\bar{z}',z')f(z',\bar{z}')\vert {_{l,\lambda}\langle z }|z'\rangle_{l,\lambda}\vert^2,
\end{equation} 
called  the lower or \emph{covariant}
  symbol of the operator $A_f$ \cite{csfks,berezin75},
the map $f \mapsto \check{f}$ is an integral transform  with the kernel
 $|{_{l,\lambda}} \langle z |{z'}\rangle_{l,\lambda}|^2 $
 which generalizes the Berezin transform.
For $z= re^{i\theta}$ and $f(z,\bar{z})=F(\theta)$,  one can find the lower (contravariant) symbol of a function $F(\theta)$ as follows
\begin{equation}
\label{fctangleop2}
 \check{f}(z,\bar z)={_{l,\lambda}}\langle z| A_f |z\rangle _{l,\lambda}
=  c_0(F) +\sum_{n\neq 0} \frac{q^{({}^n_2)}}{(q;q)_n}\Big(\frac{(1-q)qr^2}{l^2q^\lambda}\Big)^n\mathcal{S}_{n}(r;F,q),
\end{equation}
 where the function $\mathcal{S}_{n}(r;F,q)$ is given by
\begin{equation}
\label{dkr}
\mathcal{S}_{n}(r;F,q)= \frac{1}{\mathcal{N}_{l,\lambda}(r^2)}\sum_{k=0}^n (-1)^{-\frac{k}{2}}(q^{-n};q)_{\frac{k}{2}}\Bigg(\frac{lq^{\lambda/2} e^{i\theta}}{\sqrt{1-q}\,r}\Bigg)^kc_k(F),
\end{equation}
with the conditions $c_0(F):= c_k(e^{ik\theta})=1.$
%

\section{Conclusion}
In this paper, 
 we have provided an explicit construction, including the recursion relation of generalized continuous  
$(q;l,\lambda)-$Hermite polynomials 
generated by polynomial expansion of  the deformed position and momentum operators in associated Fock space basis.  Particular classes of deformed Hermite polynomials have been deduced  with explicit  weight functions. 
The diagonal representation of the density matrix using the $(q;l,\lambda)-$CS has been computed. Reproducing kernel $K(z,\zeta)$  and its properties have been  investigated. Main matrix elements of normal and anti normal forms  and  mean operator values  have been determined.  The Berezin-Klauder-Toeplitz  quantization (also called CS quantization) of classical phase space observables has been performed.  Furthermore, the  angle and   time evolution  operators and  semi-classical phase space trajectories have been  discussed.

\section*{Acknowledgements}
This work is partially supported by the Abdus Salam International
Centre for Theoretical Physics (ICTP, Trieste, Italy) through the
Office of External Activities (OEA) - \mbox{Prj-15}. The ICMPA
is also in partnership with
the Daniel Iagolnitzer Foundation (DIF), France.

\section*{Appendix A. \;{Hopf algebra structure of the 
$(q;l,\lambda)-$deformed oscillator algebra}}
In this appendix, we show that the generalized 
Heisenberg-Weyl algebra, generated
by the generators $N, \,a, \, a^\dag$ and the 
relations (\ref{rela}) carries out a Hopf algebra structure.

An algebra $\mathcal{C}$ is a Hopf algebra if 
there are defined a coproduct $\Delta,$ a counit 
$\epsilon$ and an anti-homomorphism
of antipode $S$ such that 
\begin{eqnarray}
\label{ant}
\Delta: \mathcal{H}\longrightarrow 
\mathcal{H}\otimes\mathcal{H},\quad \Delta(AB)=\Delta(A)\Delta(B)\\
\label{anti}
\epsilon: \mathcal{H}\longrightarrow 
\mathcal{H},\quad \epsilon (AB)=\epsilon(A)\epsilon(B),\\
S(AB)=S(A)S(B),
\end{eqnarray}
which satisfy the properties
\begin{eqnarray}
\label{a1}
(id\otimes \Delta)\Delta(h)=(\Delta\otimes id )\Delta(h)\\
\label{a2}
(id\otimes \epsilon)\Delta(h)=(\epsilon\otimes id )\Delta(h)\\
\label{a3}
m(id\otimes S)\Delta(h)=m(S\otimes id )\Delta(h)=\epsilon(h)I,
\end{eqnarray}
for all $h\in\mathcal{H}$ and $m$ is the multiplication 
defined as $m: \mathcal{H}\longrightarrow 
\mathcal{H}\otimes\mathcal{H}.$
To prove this it is sufficient to show that these 
relations are satisfied by the generators governing the considered algebra.
Using the Leibniz rule, we have
\begin{eqnarray}
\label{aasaa}
{_q}\partial_x^{l,\lambda}(fg)(x)={_q}\partial_x^{l,\lambda}(f(x))g(x)+{_q}\tau_x^{-1}f(x){_q}\partial_x^{l,\lambda} g(x),
\end{eqnarray}
for $f,g\in\mathcal{O}(\mathbf{D}_{l,\lambda}),$ the set of holomorphic 
functions defined on the disc $\mathbf{D}_{l,\lambda}.$
Let the actions of coproduct $\Delta$, counit $\epsilon$, and 
antipode $S$ on the generators of the algebra be defined as 
follows 
\begin{eqnarray}
\label{del}
\Delta(a^\dag)=c_1l^2q^\lambda a^\dag\otimes q^{\alpha_1 N}+c_2l^2q^\lambda q^{\alpha_2 N}\otimes a^\dag,\\
\label{aqq}
\Delta(a)=c_3l^2q^\lambda a\otimes q^{\alpha_3 N}+c_4l^2q^\lambda q^{\alpha_4 N}\otimes a,\\
\label{deltan}
\Delta(N)=c_5N\otimes I+c_6I\otimes N+\gamma I\otimes I, \quad \Delta{I}=I\otimes I,\\
\label{detas}
\epsilon(a^\dag)=c_7,\quad \epsilon(a)=c_8,\quad \epsilon(N)=c_9,\quad  \epsilon(I)=I,\\
\label{aaessss}
S(a^\dag)=-c_{10}a^\dag, \;S(a)=c_{11}a, \; S(N)=-c_{12}N+c_{13}I,\; S(I)=I,
\end{eqnarray}
where the constants $c_k, k=1,2,\ldots, 13$,  $\alpha_j=1,2,\ldots,4$ and $\gamma$ are unknown coefficients 
depending on the
Hopf algebra properties. Then 
\begin{enumerate}
\item From (\ref{a1}) and  (\ref{del}),  for $h=a^\dag$ we find
\begin{eqnarray}
\label{usinge}
c_1=l^{-2}q^{\alpha_1\gamma-\lambda},\quad c_2=l^{-2}q^{\alpha_2\gamma-\lambda},\quad c_5=1=c_6.
\end{eqnarray}
\item From (\ref{a1}) and (\ref{aqq}),  for $h=a$ we find
\begin{eqnarray}
\label{uesinge}
c_3=l^{-2}q^{\alpha_3\gamma-\lambda},\quad c_4=l^{-2}q^{\alpha_4\gamma-\lambda},\quad c_5=1=c_6.
\end{eqnarray}
\item A direct computation gives 
\begin{eqnarray}
\label{ueesinge}
\Delta ([a,a^\dag]_\alpha)&=&(l^2q^\lambda)^2\Big(c_1c_3[a,a^\dag]_\alpha\otimes q^{(\alpha_1+\alpha_3)N}+c_2c_4q^{(\alpha_2+\alpha_4)N}\otimes [a,a^\dag]_\alpha\cr
&=& c_2c_3(1-\alpha q^{-\alpha_2-\alpha_3})aq^{\alpha_2N}\otimes q^{\alpha_3N}a^\dag+
c_1c_4(1-\alpha q^{-\alpha_1-\alpha_4})\cr
&\times&q^{\alpha_4N}a^\dag\otimes q^{\alpha_1N}a\Big).
\end{eqnarray}
Therefore, $
[a,a^\dag]_\alpha=aa^\dag-\alpha a^\dag a=\frac{l^2q^\lambda}{q-1}(1-\alpha-q^{-1}(1-q\alpha)q^{-N})
$
implies that
\begin{eqnarray}
\label{usa}
\Delta ([a,a^\dag]_\alpha)=\frac{l^2q^\lambda}{q-1}(1-\alpha-q^{-1-\gamma}(1-q\alpha)q^{-N}\otimes q^{-N}).
\end{eqnarray}
Comparing (\ref{ueesinge}) and (\ref{usa}) and 
setting $
1-\alpha q^{-\alpha_2-\alpha_3}=0= 1-\alpha q^{-\alpha_1-\alpha_4}, 
$
we have
\begin{eqnarray}
  \alpha_2+\alpha_3= \alpha_1+\alpha_4,\quad \alpha_1+\alpha_3= \alpha_2+\alpha_4=-1,\quad q^{-\gamma}=2.
\end{eqnarray}
Therefore, $
 \alpha_1=\alpha_2 = \alpha_3=\alpha_4=-\frac{1}{2}.$
\item From (\ref{rela}) and (\ref{detas}), we find that 
\begin{eqnarray}
c_7=c_8=0.
\end{eqnarray}
\item From the identity (\ref{a3}) and for $h=a^\dag, a$ respectively, we get
\begin{eqnarray}
c_{12}=-1,\quad c_{10}=-q^{-\frac{c_{13}}{2}}=-c_{11}.
\end{eqnarray}
\end{enumerate}
Finally the following definition of Hopf algebra structure is in order
\begin{eqnarray}
\Delta(a^\dag)=q^{-\frac{\gamma}{2}}( a^\dag\otimes q^{-\frac{ N}{2}}+ q^{-\frac{ N}{2}}\otimes a^\dag),\\
\Delta(a)= q^{-\frac{\gamma}{2}}(a\otimes q^{-\frac{ N}{2}}+ q^{-\frac{ N}{2}}\otimes a),\\
\Delta(N)=N\otimes I+I\otimes N+\gamma I\otimes I, \quad \Delta{I}=I\otimes I,\\
\epsilon(a^\dag)=0= \epsilon(a),\quad \epsilon(N)=-\gamma,\quad  \epsilon(I)=I,\\
S(a^\dag)=q^{-\frac{c_{13}}{2}}a^\dag, \; S(a)=q^{-\frac{c_{13}}{2}}a, \;S(N)=N+c_{13}I,\;S(I)=I.
\end{eqnarray}
Indeed, for $h=N,$ we have
\begin{eqnarray}
m(S\otimes id)\Delta(h)=2N+\gamma I+ c_{13}=m(id\otimes S)\Delta(h),
\end{eqnarray}
and
\begin{eqnarray}
(id\otimes \epsilon)(\Delta(N))=(\epsilon\otimes id )(\Delta(N)).
\end{eqnarray}

\section*{Appendix B.}
From   (\ref{ket}), a direct computation gives
\begin{eqnarray}
\label{eele}
\fl
\langle r|a^{n}a^{\dag m}|s\rangle
&=&\frac{q^{\frac{1}{2}({}^r_2)+\frac{1}{2}({}^s_ 2)}}{\sqrt{\gamma^{r+s}(q;q)_r(q;q)_s}}\langle 0|a^{r+n}a^{\dag m+s} |0\rangle\cr
&=&q^{-\frac{1}{2}({}^n_2)-\frac{1}{2}({}^{m}_{\, 2})-\frac{rn+sm}{2}}
\sqrt{\gamma^{m+n}(q^{1+r};q)_{n}(q^{1+s};q)_{m}}\;\delta_{r+n,s+m},
\end{eqnarray}
for any integers $n$ and $m$.\\
Let us make use of the following results to show (\ref{eees}) and (\ref{scend})
\begin{eqnarray}
\label{eaaaa}
a^{\dag m}a^n&=&a^{\dag m-1} a^{n-1}\varphi(N-n+1)\cr
&=&a^{\dag m-2}a^{n-2}\varphi(N-n+2)\varphi(N-n+1)\cr
&\vdots&\cr
&=&a^{\dag m-n} \prod_{k=0}^{n-1}\varphi(N-k),\quad \mbox{ for }\quad  n< m.
\end{eqnarray}
Similarly, 
\begin{eqnarray}
\label{tddd}
a^{\dag m}a^n
=  \prod_{k=0}^{m-1}\varphi(N-k)a^{n-m}, \quad \mbox{ for } \quad n>m.
\end{eqnarray}
From (\ref{eaaaa}) and (\ref{tddd}), a direct computation gives 
\begin{eqnarray}
\fl
\langle r|a^{\dag m}a^{n}|s\rangle&=&\frac{q^{\frac{1}{2}({}^r_2)+\frac{1}{2}({}^s_ 2)}}{\sqrt{\gamma^{r+s}(q;q)_r(q;q)_s}}\langle 0|a^{r}a^{\dag m}A^nA^{\dag s}|0\rangle\cr
&=&(- q)^n(q^{-s};q)_nq^{-\frac{1}{2}({}^{m-n}_{\;\;\;2})-\frac{s(m-n)}{2}}
\sqrt{\gamma^{m+n}(q^{1+s};q)_{m-n}}\;\delta_{r,m-n+s},
\end{eqnarray}
if $n<m$,
\begin{eqnarray}
\fl
\langle r|a^{\dag m}a^{n}|s\rangle&=&\frac{q^{\frac{1}{2}({}^r_2)+\frac{1}{2}({}^s_ 2)}}{\sqrt{\gamma^{r+s}(q;q)_r(q;q)_s}}\langle 0|a^{r}a^{\dag m}a^na^{\dag s}|0\rangle\cr
&=&(- q)^m(q^{n-s-m};q)_m q^{-\frac{1}{2}({}^{n-m}_{\;\;\;2})-\frac{s(n-m)}{2}}
\sqrt{\gamma^{m+n}(q^{1+r};q)_{n-m}}\;\delta_{s,n-m+r},\cr
&&
\end{eqnarray}
if $n>m.$


\section*{References}
\begin{thebibliography}{32}
\bibitem{Akhiezer} Akhiezer N I 1965 {\it The Classical Moment Problem and Some Related Questions in Analysis}, Olivier and Boyd, London 
\bibitem{arik}
 Arik M and Coon D D 1976
  {\rm Hilbert Spaces of Analytic Functions and Generalized Coherent States},
  {\it J. Math. Phys.} {\bf 17}, 524
\bibitem{acc} Aronszajn N 1950 {\rm Theory of reproducing kernels}, {\it Trans. Am. Math. Soc.}, 
{\bf  68},  337-404
\bibitem{askey}
Askey R 1989 {\it Continuous $q-$Hermite polynomials 
when $q>1$}, {\it in $q-$Series and Partitions,} edited by 
Stanton D (Springer-Verlag, New York)
\bibitem{Atakishiyeva} Atakishiyev N M and   
Atakishiyeva M K  2001 {\rm A $q$-analogue 
of the Euler gamma integral},  {\it Theor. Math. Phys.} {\bf 129}, 1325-1334

\bibitem{Balo}Balo\"itcha E,  Hounkonnou M N  and Ngompe Nkouankam E B 2012 {\rm
$(p, q;\alpha,\beta,  \nu, \gamma)-$deformed oscillator algebra: Irreducible 
representations and induced deformed harmonic oscillator},
 {\it J. Math. Phys.} {\bf 53} 013504
\bibitem{berezin75} 
Berezin F A 1975  { \rm Commun. Math. Phys}. \textbf{40}, 153--174 
\bibitem{Bergeron}  Bergeron H, Siegl P and Youssouf A  2012 {\rm 
New SUSYQM coherent states for P\"oschl - Teller 
potentials: a detailed mathematical analysis},  {\it J. Phys. A: Math. Theo.}  {\bf 45} 
244028
\bibitem{dez} Bukweli Kyemba J D   and Hounkonnou 
M N 2012 {\rm $(q;l,\lambda)-$deformed Heisenberg 
algebra: coherent states, their statistics and 
geometry},  Afr. Diaspora J. Math, {\bf 14}  pp 38 - 56
\bibitem{chiu} Chiu S H, Gray R W  and Wilson C A 1992  {\it J. Phys. Lett. A} {\bf 164}  237 
\bibitem{Dancoff} Dancoff S M 1950 {\rm  Non-adiabatic 
meson theory of nuclear forces} {\it Phys. Rev.} {\bf 78} 382-385
\bibitem{ext} Exton M {\it $q-$Hypergeometric Functions and 
Applications}, {\it Ellis Harwood Limited}
\bibitem{gazeau09} 
Gazeau J P 2009 {\it Coherent States in Quantum Physics}
  (Wiley-VCH: Berlin)
\bibitem{gazeautt} Gazeau J P and  del Olmo M A 
2012 {\rm Pisot $q$-Coherent states quantization 
of the harmonic oscillator} {\it arxiv: 1207.1200}
\bibitem{glauber} Glauber R J 1963 {\rm  Coherent and incoherent states of 
radiation field}, {\it Phys. Rev.}, V. 134, (6),  2766-2788 
\bibitem{klauder95} 
Klauder J  R  1995 
 {\rm Ann. of Phys}.  {\bf 237},  147-160  
\bibitem{klauder63}
 Klauder J R 1963   
{ \it J. Math. Phys.}  
 {\bf 4}, 1055--1058 and  1058--1073
\bibitem{ASK}  Koekoek R  and Swarttouw R F 1998
{\it The Askey-scheme of orthogonal 
polynomials and its $q-$analogue},
Report 98 -17, TU Delft
\bibitem{csfks}
Lieb E 1994 {\it Coherent States: Past, Present and Future}  p. 267--278
%
\bibitem{ahi} Maximov V and Odzijewicz A 1995  {\rm The $q-$deformation
of quantum mechanics of one degree of 
freedom},  {\it J. Math. Phys}, {\bf 36} http://dx.doi.org/10.1063/1.531080
\bibitem{part1} Parthasarathy R  and Sridhar R 2002 {\rm A diagonal 
representation of the quantum density matrix using $q-$boson 
oscillator coherent  states},  {\it Phys. Lett. A}, {\bf 305} 105-110
\bibitem{part2} Parthasarathy R and Sridhar R 
2004  {\rm Diagonal representation of density 
matrix using $q-$coherent  states},  {\it Proc. of Institue of 
Math.  NAS of Ukraine}, {\bf 50}  Part 2, 909-914
\bibitem{Perelomov:1995ye} 
  Perelomov A M  1995
  {\rm On the Completeness of some subsystems 
of q-deformed coherent states,}
  {\it Helv. Phys. Acta} {\bf 68}, 554 
\bibitem{quesne} Quesne C 2002 {\rm New q-deformed coherent states 
with an explicitly known resolution of 
unity, } {\it J. Phys. A: Math. Gen}  {\bf  35}, no. 43, pp. 9213-9226
\bibitem{reed} Reed F and Simon B 1975 {\it Methods of Modern 
Mathematical Physics:  Fourier Analysis, Self-Adjointness} (New York: Academic)
\bibitem{Snyder} Snyder H S 1947 {\rm Quantized space-time,} {\it Phys. Rev. } {\bf 71} 38-41
\bibitem{sudar} Sudarshan E C G 1963 {\rm Equivalence of 
semi-classical and quantum mechanical descriptions of statistical light
beams}, {\it Phys. Rev. Lett.},  {\bf  10}, (7), 277-279
\bibitem{Tarmakin} Tamarkin J D  and Shoha J A 
1943 {\it The Problem of Moments}, A. P. S., New York
\bibitem{Tamm} Tamm I E  1945 {\rm The relativistic 
interaction of elementary particles}  {\it J. Phys. UdSSR} {\bf 9} 449-465
\bibitem{alit}  Twareque Ali S, Fabio Bagarello and Jean Pierre 
Gazeau 2012 {\rm Quantizations from reproducing kernel 
spaces } arxiv 1212.3664
\bibitem{sang} Won-Sang Chung and Klimyk A U 1996 {\rm On 
position and momentum operators
in the $q-$oscillator algebra,} {\it J. Math. Phys.} {\bf 37 } 531419 
\bibitem{zhang}
Zhang W-M, Feng D H and Gilmore R 1990 {\rm Coherent 
states: Theory and some applications,} {\it Reviews 
of Modern Physics.} {\bf 62}\ 887--927  


\end{thebibliography}
\end{document}